\documentclass[12pt]{article}
\usepackage{epsf,latexsym}
\usepackage[all]{xy}
\usepackage{amsfonts,amssymb} 
\epsfverbosetrue
\newcommand{\double}[1]{\mathbb{#1}}

\newcommand{\Cc}{\double{C}}

\newcommand{\Rr}{\double{R}}

\newcommand{\D}{\mathcal{D}}

\newcommand{\lb}{\left[}
\newcommand{\rb}{\right]}

\newcommand{\bb}{\begin{eqnarray}}
\newcommand{\ee}{\end{eqnarray}}
\newcommand{\eee}{\nonumber\end{eqnarray}}

\begin{document}

\font\twelve=cmbx10 at 13pt
\font\eightrm=cmr8

\thispagestyle{empty}

\begin{center}
${}$
\vspace{3cm}

{\Large\textbf{Ashtekar's variables without spin}} \\

\vspace{2cm}

{\large Thomas Sch\"ucker 

\vspace{2cm}
\centerline{Institut f\"ur Theoretische Physik}
\centerline{der Universit\"at Heidelberg}
\centerline{Philosophenweg 16, D-6900 Heidelberg}
}

\vspace{2cm}

\vskip 2cm

{\large\textbf{Abstract}}
\end{center}
Ashtekar's variables are shown to arise naturally from a $3+1$
split of general relativity in the Einstein-Cartan formulation.
Thereby spinors are exorcised.\vspace{2cm}

\vskip 1truecm

\noindent HD-THEP-88-12\\
\noindent May 1988\\
\vspace{1cm}

${}$

Ashtekar's variables \cite{1} have already received much
attention \cite{2} and need no introduction. It is intended to
show that they arise naturally from the Einstein-Cartan formulation
of general relativity without the use of spinors. We follow the
notations of reference \cite{3} where a detailed presentation
of the Einstein-Cartan theory may be found.
\smallskip
Cartan's key idea is to describe a metric by an orthonormal
frame of vector fields $e_a,\ a=0,1,2,3$:
\bb g(e_a,e_b)=\eta_{ab}\ee
\bb (\eta_{ab})=diag(+1,-1,-1,-1)\ee
or its dual frame of 1-forms $e^a,\ a=0,1,2,3:$
 \bb e^a(e_b)=\delta^a_{\phantom{a}b}.\ee
Of course two frames $e_a$ and ${e'}_a$ related by a Lorentz
transformation $\Lambda\in SO(1,3)$
 \bb {e'}_a=\Lambda^{-1b}_{\phantom{-1b}a}e_b,\ee
 \bb \Lambda^T\eta\Lambda=\eta\ee
describe the same metric. In order to construct an invariant action
one introduces a connection $\omega$, i.e. a 1-form on space-time
$M$ with values in the Lie algebra $so(1,3)$, which then preserves
the metric under parallel transport. Under a change of frame (4)
the connection transforms as
\bb \omega'=\Lambda\omega\Lambda^{-1}+\Lambda d\Lambda^{-1}.\ee
In these variables $e$ and $\omega$ the Einstein-Hilbert action reads:
 \bb S_{EH}(e,\omega):={{-1}\over{32\pi G}}\int_MR^a_{\phantom{a}
 b'}\eta
^{b'b}\wedge e^c\wedge e^d\epsilon_{abcd}\ee
where $R$ is the curvature of $\omega$, a 2-form with values
in $so(1,3)$:
 \bb R=d\omega+{1\over2}\lb\omega,\omega\rb\ee
and $\epsilon_{abcd}$ is completely antisymmetric with
$\epsilon_{0123}=1$.
Variations with respect to the (dual) frame give the Einstein
equations
 \bb R^{ab}\wedge e^d\epsilon_{abcd}=0\ee
while varying the connection and partial integration yields
vanishing torsion
 \bb T^a:=De^a:=de^a+\omega^a_{\phantom{a}b}\wedge e^b=0.\ee
These linear equations in $\omega$ can be solved uniquely to
express the connection as a function of the frame components
and their first derivatives, the so-called Riemannian
connection.
\smallskip
Choose a time coordinate $t:M\to\Rr$
 \bb\buildrel \star \over g (dt,dt)>0\ee
 $\buildrel \star \over g$ being the induced metric for
1-forms. A 3+1 split is a parametrization of all metrics on
$M$ by starting from a parametrization of all metrics on the
3-surfaces $\sum_t$ of ``simultaneity''. In the frame formulation
this imposes the use of vector fields rather than 1-forms.
\smallskip
Let $e_{\bar a}$, $\bar a=1,2,3$ be three linearly independent
vector fields tangent to $\sum_t$:
 \bb dt(e_{\bar a})=0,\quad \bar a=1,2,3.\ee
 Modulo $SO(3)$ rotations they parametrize the metrics on $\sum_t$.
This parametrization is completed to include all metrics on $M$
by a vector field $e_0$ pointing towards the future:
 \bb dt(e_0)>0.\ee
In coordinates $x^0=t,\quad x^{\bar \mu},\quad \bar\mu=1,2,3$
we have
 \bb e_a=(\gamma^{-1})^\mu_{\phantom{\mu}a}
 {\partial\over{\partial x^\mu}},\ee
\bb (\gamma^{-1})^0_{\phantom{0}\bar a}=0\qquad \bar a=1,2,3.
\ee
Consequently there are 1+3+9=13
variables $(\gamma^{-1})^0_{\phantom
{0}0},\ (\gamma^{-1})^{\bar\mu}_{\phantom{\bar\mu}0},\ (\gamma
^{-1})^{\bar\mu}_{\phantom{\mu}\bar a}$. From the 9 $(\gamma^{-1})^
{\bar\mu}_{\phantom{\mu}\bar a}$ we still have to subtract 3
degrees of freedom for the $SO(3)$ transformations and finally
we remain with 10 variables parametrizing all metrics. This
parametrization was introduced in 1963 by Schwinger \cite{4}
under the name ``time gauge''.
\smallskip
For later convenience we rename the components of the frame
 \bb dx^\mu(e_a)=\left(\gamma^{-1}\right)^\mu_{\phantom\mu
 a}=:8\pi G\ det\left(
\pi^{\bar\nu}_{\phantom{\bar\nu}\bar b}\right)\pmatrix{
{1\over N}&0\cr
{{n^{\bar\mu}}\over N}& \pi^{\bar\mu}_{\phantom {\mu}\bar a}\cr}.
\ee
Let
\bb\omega^a_{\phantom{a}b}=:\omega^a_{\phantom {a}b\mu}dx^\mu
\ee
be the components of the connection. From now on we raise Latin
(Lorentz-)indices with $\eta^{ab}=\eta_{ab}$. A bar over a Latin
or Greek (coordinate) index indicates that it only takes spatial
values. We can then write the Einstein-Hilbert action as:
\bb S_{EH}(e,\omega)&=&\int^{+\infty}
 _{-\infty}dt\int_{\sum_t}
dx^1dx^2dx^3\left[ 
\left(n^{\bar\nu}\pi^{\bar\mu}_{\phantom{\mu}\bar b}-n^{\bar\mu}
\pi^{\bar\nu}_{\phantom{\nu}\bar b}\right)\left(\partial_{\bar\mu}
\omega_{\phantom{0b}\bar\nu}
^{0\bar b}-\omega^{0\bar c}_{\phantom{0c}\bar\mu}\omega^{\bar c
\bar b}_{\phantom{cb}\bar\nu}\right)\right.\nonumber\\&&
-{1\over2}N\left(\pi^{\bar\mu}_{\phantom {\mu}
\bar a}\pi^{\bar\nu}_{\phantom{\nu}\bar b}-\pi
^{\bar\nu}_{\phantom{\nu}
\bar a}\pi^{\bar\mu}_{\phantom{\mu}\bar b}\right)
\left(\partial
_{\bar\mu}\omega^{\bar a\bar b}_{\phantom{ab}
\bar\nu}-\omega^{\bar a\bar c}_{\phantom{ac}\bar
\mu}\omega^{\bar c\bar b}_{\phantom{cb}
\bar\nu}-\omega^{0\bar a}_{\phantom{0b}\bar\mu}
\omega^{0\bar b}_{\phantom{0b}\bar\nu}\right)\nonumber\\&&\left.
-\pi^{\bar\nu}_{\phantom{\nu}
\bar b}\Bigl(\partial_0\omega^{0\bar b}_{\phantom{0b}\bar\nu}
-\partial_{\bar\nu}\omega^{0\bar b}_{\phantom{0b}
0}- \omega^{0\bar c}_{\phantom{0c}0}\omega
^{\bar c\bar b}_{\phantom{cb}
\bar\nu}+\omega^{0\bar c}_{\phantom{0c}
\bar\nu}\omega^{\bar c\bar
b}_{\phantom{cb}0}\Bigr)\right]. \ee
In the variables $N,\  n^{\bar\mu},\ \pi^{\bar\mu}
_{\phantom{\mu}a}$ and
$\omega^{ab}_{\phantom{ab}
\mu}$ the action is \underbar{polynomial}. Furthermore
the requirement that the three vector fields $e_{\bar a}$ be
linearly independent may be
dropped allowing also \underbar{degenerate}
$\pi^{\bar\mu}_{\phantom{\mu}
\bar a}$. The only dynamical variables are the
nine functions $\omega ^{0\bar b}_{\phantom{0b}
\bar\nu}$ which are related
to the extrinsic curvature of $\sum_t$. Their time derivative only
appears linearly in the action and their conjugate momenta are the
nine $-\pi^{\bar\mu}_{\phantom{\mu}\bar a}$. Variation with respect to
$N,\ n^{\bar\mu}$ and $ \pi^{\bar \mu}_{\phantom{\mu}
\bar a}$
 yields the Einstein equations, while varying
$\omega$ and a partial integration gives again the torsion
zero equation in the following form:
 \bb\delta\omega^{0\bar b}_{\phantom{0b}0}:\qquad
D_{\bar\nu}\pi^{\bar\nu}_{\phantom{\nu}
\bar b}:=\partial_{\bar\nu}\pi^{\bar\nu}
_{\phantom{\nu}\bar b}+\omega_{\bar b
\phantom{\bar c}\bar \nu}^{\phantom{\bar b}
\bar c}\pi^{\bar\nu}_{\phantom{\nu}\bar c}=0 .
\label{19}\ee
$D_\mu$ is the $SO(3)$-covariant derivative with respect to
$\omega^{\bar a}_{\phantom{a}\bar b\mu}$.
 \bb\delta\omega^{\bar a\bar b}_{\phantom{\bar a\bar b}0}:
\qquad \omega^0_{\phantom{0}
\bar a\bar \nu}\pi^{\bar\nu}_{\phantom{\nu}\bar b}-\omega
^0_{\phantom{0}\bar b\bar \nu}\pi^{\bar\nu}_{\phantom{\nu}
\bar a}=0, \label{20}\ee
 \bb&\delta\omega^{\bar a\bar b}_{\phantom{ab}
 \bar\mu}:\quad
(\omega^0_{\phantom{0}
\bar a 0}\pi^{\bar\mu}_{\phantom{\mu}\bar b}-\omega^0_{\phantom{0}
\bar b 0}\pi^{\bar\mu}_{\phantom{\mu}\bar a})
+\omega^0_{\phantom{0}
\bar a\bar\nu}(n^{\bar\nu}\pi^{\bar \mu}_{\phantom{\mu}
\bar b}-n^{\bar\mu}\pi^{\bar\nu}_{\phantom{\nu}
\bar b})-\omega^0_{\phantom{0}\bar b\bar\nu}(n^{\bar\nu}\pi^{\bar
\mu}_{\phantom{\mu}\bar a}-n^{\bar\mu}\pi^{\bar\nu}
_{\phantom{\nu}\bar a})&\nonumber\\&
-D_{\bar\nu}(N\pi^{\bar\nu}_{\phantom{\nu}\bar a}
\pi^{\bar\mu}_{\phantom{\mu}\bar b}
-N\pi^{\bar\nu}_{\phantom{\nu}
\bar b}\pi^{\bar\mu}_{\phantom{\mu}\bar a})=0,\label{21}&\ee
\bb
 \delta\omega^{0\bar b}_{\phantom{0b}
\bar\mu}:\qquad D_0\pi^{\bar\mu}_{\phantom{\mu}\bar b}
+D_{\bar\nu}(n^{\bar\nu}\pi^{\bar\mu}_{\phantom{\mu}
\bar b}-n^{\bar\mu}\pi^
{\bar\nu}_{\phantom{\mu}\bar b})
+\omega^{0\bar c}_{\phantom{0c}
\bar\nu}(N\pi^{\bar\nu}_{\phantom{\nu}\bar c}\pi^{\bar\mu}
_{\phantom{\mu}
\bar b}-N\pi^{\bar\mu}_{\phantom{\mu}
\bar c}\pi^{\bar\nu}_{\phantom{\nu}\bar b})=0.
\label{22}\ee
Now, if we replace the 6 real-valued 1-forms $\omega^{ab}$ by
the 3 complex-valued 1-forms $\chi^{\bar a\bar b}$
\bb\chi^{\bar a\bar b}:=\omega^{\bar a\bar b}+i\epsilon
^{\bar a\bar b}_{\phantom{ab}
0\bar c}\omega^{0\bar c}\ee
and if we denote by $\D_\mu$ the covariant
derivative with respect to $\chi^{\bar a\bar b}_{\phantom{ab}
\mu}$, then the
6 real equations (\ref{19}) and (\ref{20}) combine into 3 complex equations:
\bb\D_{\bar\nu}\pi^{\bar\nu}_{\phantom{\nu}\bar b}=0.\label{24}\ee
Likewise,  (21) and (22) condense to
\bb\D_0\pi^{\bar\mu}_{\phantom{\mu}
 \bar b}+\D_{\bar\nu}(n^{\bar\nu}\pi^{\bar\mu}
_{\phantom{\mu}
\bar b}-n^{\bar\mu}\pi^{\bar\nu}_{\phantom{\nu}
\bar b})+i\epsilon_{0\bar b}
^{\phantom{0b}
\bar c\bar d}\D_{\bar\nu}(N\pi^{\bar\nu}
_{\phantom{\nu}\bar c}\pi^{\bar\mu}_{\phantom{\mu}\bar d})
=0.\label{25}\ee
Samuel $\lb5\rb$ and Jacobson and Smolin $\lb6\rb$ have remarked
that these complex equations can conveniently be obtained by
adding to the Einstein-Hilbert action a piece that vanishes
``on shell'':
\smallskip
Let
 \bb\phi:={1\over2}(\omega+i\tilde\omega)\ee
where $\tilde\omega$ is the dual connection:
 \bb{\tilde\omega}^{ab}:={1\over2}\epsilon^{ab}_{\phantom{ab}
 cd}\omega^{cd}.
\ee
Note the difference between dual and Hodge star. $\phi$ is complex
but self-dual in the sense that
 \bb\tilde\phi=-i\phi.\ee
Consequently
 \bb\phi^{\bar a\bar b}=i\epsilon^{\bar a\bar b}_{\phantom{ab}
 0\bar c}\phi
^{0\bar c}.\ee
Replacing the real $\omega$ by the complex
$\phi$ in the Einstein-Hilbert
action amounts to adding to the real Einstein-Hilbert action a purely
imaginary piece:
 \bb S_{EH}(e,\phi)={1\over2}S_{EH}(e,\omega)+{i\over 2}S_I(e,\omega)
\ee
where
 \bb S_I(e,\omega):={{-1}\over{32\pi G}}\int_M \tilde R^{ab}\wedge
e^c\wedge e^d\epsilon_{abcd}={{-1}\over{16\pi G}}\int_M R_{cd}
\wedge e^c\wedge e^d.\ee
Indeed the curvature of $\phi$ is
 \bb d\phi+{1\over2}\lb\phi,\phi\rb={1\over2}R+{i\over 2}\tilde R
\ee
where $\tilde R$ is the dual of the curvature $R$ of $\omega$.
\smallskip
The field equations of the complex action $S_{EH}+iS_I$ are the
same as the field equations of the real Einstein-Hilbert action
alone, because varying $S_I$ with respect to the frame yields
the Bianchi identity:
 \bb 0=DT=R\wedge e\ee
and variation with respect to the connection gives again vanishing
torsion.
\smallskip
We define the connection $\chi$, a 1-form on $M$ with values in
$so(3,\Cc)$ by
\bb\chi^{\bar a\bar b}:=2\phi^{\bar a\bar b}=
\phi^{\bar a\bar b}+i\epsilon^{\bar a\bar b}_{\phantom{ab}0\bar c}
\phi^{0\bar c}
=\omega^{\bar a\bar b}+i\epsilon^{\bar a\bar b}_{\phantom{ab}0\bar c}
\omega^{0\bar c}\ee
and denote by $\D$ its covariant derivative and by $F$
its curvature
 \bb F^{\bar a}_{\phantom{a}\bar b}:=
 d\chi^{\bar a}_{\phantom{a}\bar b}+\chi^{\bar a}
_{\phantom{a}\bar c}\chi^{\bar c}
_{\phantom{c}\bar b}=:{1\over 2}F^{\bar a}
_{\phantom{a}\bar b\mu\nu}
dx^\mu\wedge dx^\nu.\ee
Then, up to a surface term from the last expression on the
right-hand side of equation (36),  the action can be written:
 \bb S_{EH}(e,\omega)+iS_I(e,\omega)&=&
 \nonumber\\&&
\int^{+\infty}_{-\infty
}dt\int_{\sum_t}d^3x\lb{i\over2}n^{\bar\nu}
\pi^{\bar\mu}_{\phantom{\mu}
\bar a}\epsilon^{0\bar a}_{\phantom{0a}\bar c\bar d}F^{\bar c\bar d}
_{\phantom{cd}\bar\mu\bar\nu}
-{1\over2}N\pi^{\bar\mu}_{\phantom{\mu}
\bar a}\pi^{\bar\nu}_{\phantom{\nu}\bar b}F^{\bar a
\bar b}_{\phantom{ab}\bar\mu\bar\nu}\right.\nonumber\\
&&\left. -{i\over 2}\epsilon^{0\bar a}_{\phantom{0a} 
\bar c\bar d}\partial_0\chi^{\bar c\bar d}
_{\phantom{cd}\bar\nu}\pi^{\bar\nu}_{\phantom{\nu}\bar a}
-{i\over 2}\epsilon^{0\bar a}_{\phantom{0a}\bar c
\bar d}\chi^{\bar c\bar d}_{\phantom{cd}
0}\D_{\bar\nu}\pi^{\bar\nu}_{\phantom{\nu}\bar a}
\rb\label{36}\ee
Variations with respect to $\chi^{\bar a\bar b}_{\phantom{ab}0}$ and
$\chi^{\bar a\bar b}_{\phantom{ab}
\bar\mu}$ yield again
equations (24) and (25), respectively. Note that
now due to the additional term $iS_I$, all 18 components of
$\chi^{\bar a\bar b}_{\phantom{ab}
\bar\mu}$ are dynamical and their momenta
are
 \bb p^{\bar\mu}_{\phantom{\mu}
 \bar a\bar b}:=-i\epsilon^{0\bar c}_{\phantom{0c}\bar a\bar b}\pi
^{\bar\mu}_{\phantom{\mu}\bar c}\ee
considered as complex variables. On the other hand, $n^{\bar\mu},
\ N$ and $\chi^{\bar c\bar d}_{\phantom{cd}0}$
 remain Lagrange multipliers.
\smallskip
Since the time derivatives of $\chi^{\bar c\bar d}
_{\phantom{cd}\bar\nu}$
appear only linearly in the action, the Hamiltonian is obtained
by deleting the third term in equation (36) and changing all signs:
 \bb H=\int_{\sum_t}d^3x\lb {1\over2}n^{\bar\nu}p^{\bar
\mu}_{\phantom{\mu}\bar c\bar d}F^{\bar c\bar d}_{\phantom{cd}
\bar\mu\bar\nu}
-{1\over8}N\epsilon^{\bar c\bar d}_{\phantom{cd}
0\bar a}p^{\bar\mu}_{\phantom{\mu}\bar c\bar d}
\epsilon^{\bar r\bar s}_{\phantom{rs}
0\bar b}p^{\bar\nu}_{\phantom{\nu}\bar r\bar s}
F^{\bar a\bar b}_{\phantom{ab}\bar\mu\bar\nu}
-{1\over 2}\chi^{\bar c\bar d}_{\phantom{cd}0}
\D_{\bar\nu}p^{\bar\nu}_{\phantom{\nu}\bar c\bar d}
\rb.\ee
\smallskip
Although the Hamiltonian is complex, its time evolution carries
real initial data $\pi,\omega$ into real data. In conclusion,
we have recovered Ashtekar's variables which in the spinless
formulation are given by $\chi^{\bar a\bar b}_{\phantom{ab}
\bar \mu}$ and
$p^{\bar\mu}_{\phantom{\mu}\bar a\bar b}$.


\begin{thebibliography}{10}
\bibitem{1}
 A. Ashtekar, Phys. Rev. Lett. {\bf 57}, 2224 (1986),
Phys. Rev. {\bf D36}, 1787 (1987)
   \bibitem{2}
 M. Mitchell Waldrop, Science {\bf 235}, 284 (1987)
     \bibitem{3}
  M. G\"ockeler, T. Sch\"ucker, {\it Differential
Geometry, Gauge Theories, and Gravity}, Cambridge University Press,
Cambridge (1987)
  \bibitem{4}
 J. Schwinger, Phys. Rev. {\bf 130}, 1253 (1963)
  \bibitem{5}
 J. Samuel, Pram$\tilde{\hbox{a}}$na, J. Phys.
{\bf 28} L429 (1987)
\bibitem{6}
 T. Jacobson, L. Smolin, Class. Quantum Grav. {\bf 5},
583 (1988)

\end{thebibliography}
\end{document}